\documentstyle[12pt,epsf]{article}
\newcommand{\be}{\begin{equation}}
\newcommand{\ee}{\end{equation}}
\newcommand{\la}{\langle}
\newcommand{\ra}{\rangle}
\newcommand{\bea}{\begin{eqnarray}}
\newcommand{\eea}{\end{eqnarray}}
\newcommand{\om}{\omega}
\newcommand{\Om}{\Omega}
\newcommand{\pa}{\partial}
\newcommand{\sig}{\hat{\sigma}}

\newcommand{\calE}{{E}}
\begin{document}
\baselineskip 2.5ex
\date{}
\title{Quantum Noise and Superluminal Propagation}
\author{Bilha Segev\\
\small{Department of Chemistry}\\
\small{Ben-Gurion University of the Negev}\\
\small{P.O.B. 653, Beer-Sheva 84105, Israel}\\
\\
Peter W. Milonni\\
\small{Theoretical Division (T-4)}\\
\small{Los Alamos National Laboratory}\\
\small{Los Alamos, New Mexico 87545}\\
\\
James F. Babb\\
\small{Institute for Theoretical Atomic and Molecular Physics}\\
\small{Harvard-Smithsonian Center for Astrophysics}\\
\small{60 Garden St.}\\
\small{Cambridge, Massachusetts 02138}\\
\\
and\\
\\
Raymond Y. Chiao\\
\small{Department of Physics}\\
\small{University of California, Berkeley}\\
\small{Berkeley, California 94720}}
\maketitle
\newpage
\centerline{Abstract}
Causal ``superluminal" effects
have recently been observed and discussed in various contexts. The question
arises whether such effects could be observed with extremely weak
pulses, and what would prevent the observation of an
``optical tachyon." Aharonov, Reznik, and Stern (ARS) [Phys. Rev. Lett.
{\bf 81}, 2190 (1998)] have argued that quantum noise will
preclude the observation of a
superluminal group velocity when the pulse consists of one or a few photons.
In this paper we reconsider this question both in a general framework and in
the specific example, suggested by Chiao, Kozhekin, and Kurizki
[Phys. Rev. Lett. {\bf 77}, 1254 (1996)], of off-resonant, short-pulse
 propagation in
an optical amplifier.
We derive in the case of the amplifier a signal-to-noise ratio that is
consistent with the general ARS conclusions when we impose their
criteria for distinguishing between superluminal propagation and
propagation at the speed $c$. However, results consistent with the
semiclassical arguments of CKK are obtained if weaker criteria are
imposed, in which case the signal can exceed the noise without
being ``exponentially large." We show that the quantum fluctuations
of the field considered by ARS are closely related to superfluorescence
noise.
More generally we consider the implications of unitarity for
superluminal propagation and quantum noise and study, in addition
to the complete and truncated wavepackets considered by ARS, the
residual wavepacket formed by their difference. This leads to the
conclusion that the noise is mostly luminal and delayed with respect
to the superluminal signal. In the limit of a very weak incident
signal pulse, the superluminal signal will be dominated by the noise part,
and the signal-to-noise ratio will therefore be very small.
\newpage\noindent
\section{Introduction}
Chiao et al. \cite{chiao},\cite{chiao1},\cite{chiao2} have shown that certain
``superluminal" effects are possible without violation of standard notions of
Einstein causality, i.e., without conveying information faster than the 
velocity
$c$ of light in vacuum. Such effects have been demonstrated experimentally in
optical tunneling \cite{steinberg},\cite{steinberg2},\cite{spielmann} and
in an electric circuit \cite{mitchell}.

It has been suggested by Chiao, Kozhekin, and Kurizki (CKK) \cite{chiao}
 that an
optical pulse can propagate superluminally in an amplifier
whose relaxation times are long compared with the pulse duration. The 
dispersion
relation they derive can be obtained directly, as follows, starting from
the formula for the refractive index of a monatomic gas:
\be
n(\om)=1+{2\pi e^2\over m}\sum_i\sum_j{N_if(i,j)\over \om_{ji}^2-\om^2}
\label{eqI1}
\ee
for $n(\om)\cong 1$, where $N_i$ is the number density of atoms in state
$i$ and $f(i,j)$ is the oscillator strength for absorption on the
$i\rightarrow j$ transition of frequency $\om_{ji}$.
Near a two--level resonance this becomes
\be
n(\om)=1+{2\pi e^2f\over m}{N_1-N_2\over \om_o^2-\om^2} \ ,
\label{eqI2}
\ee
where 1 and 2 designate the lower-- and upper--energy levels, respectively,
and $\om_o=\om_{21}$.
Close to the transition resonance frequency $\om_o$,
\be
n(\om)\cong 1+{\pi e^2f\over m\om_o}{N_1-N_2\over\om_o-\om-i\beta}
\label{eqI3}
\ee
when we include a dipole damping rate $\beta$.
The (real) refractive
index near a resonance is then
\be
n_R(\om)=1+{\pi e^2f\over
m\om_o}{\om_o-\om\over(\om_o-\om)^2+\beta^2}(N_1-N_2) \ .
\label{eqI4}
\ee
Introducing the inversion $w=(N_2-N_1)/N$, where $N$ is the number density
 of atoms,
and assuming
 a field
sufficiently far from resonance that $(\om_o-\om)^2>>\beta^2$, we have
\be
n_R(\om)\cong 1-{\pi e^2Nwf\over m\om_o}{1\over\om_o-\om} \ ,
\label{eqI5}
\ee
\be
k=n_R(\om){\om\over c}={\om\over c}\left[1-{\pi e^2Nwf\over
m\om_o}{1\over \om_o-\om}\right]={\om\over
c}\left[1-{\om_p^2w/4\om_o\over\om_o-\om}\right] \ ,
\label{eqI6}
\ee
\be
k-k_o={1\over c}(\om-\om_o)-{\om\over c}{\om_p^2w/4\om_o\over\om_o-\om}
\cong {1\over c}(\om-\om_o)-{\om_p^2w/4c\over\om_o-\om} \ ,
\label{eqI7}
\ee
and
\be
\Om^2-Kc\Om+{1\over 4}w\om_p^2=0 \ ,
\label{eqI9}
\ee
where $K=k-k_o$, $\Om=\om-\om_o$, and the ``plasma frequency" $\om_p$ is
defined by
\be
\om_p^2=4\pi Ne^2f/m=8\pi Nd^2\om_o/\hbar \ ,
\label{pf}
\ee
with $d$ the electric dipole transition moment.
Equation (\ref{eqI9})
is the dispersion relation obtained by CKK.

We refer the reader to the CKK paper for a discussion of this
dispersion relation.
Here we simply note that
(\ref{eqI7}) implies the group velocity
\be
v_g={d\om\over dk}=c\left[1-{\om_p^2w/4\over(\om_o-\om)^2}\right]^{-1} \ ,
\label{eqI10}
\ee
so that, in the case of an amplifier ($w>0$), a short off-resonant
pulse can
propagate with a group velocity
$v_g>c$. 

Questions have been raised about the validity of the
latter prediction at the one-photon
level, which would correspond to what CKK call an
``optical tachyon" \cite{chiao}.
Aharonov, Reznik, and Stern (ARS) \cite{ars} have
presented general arguments, based on the unitary evolution of the state
vector, that ``strongly questions the possibility that these systems may 
have
tachyonlike quasiparticle excitations made up of a small number of photons."
They also consider a particular model as an analog of the CKK system.

In this paper we address the question of superluminal propagation at the 
one- or
few-photon level, and in particular the role played by quantum noise in the
propagation of such extremely weak pulses. We begin in the following section
with some physical considerations about the observability of superluminal
propagation, and we briefly compare the ARS and CKK models.
In Section 3 we formulate the Heisenberg equations of motion for the 
propagation
of a short optical pulse in an inverted medium, and briefly review some
relevant results from the theory of superfluorescence (SF). In Section
4 we derive a signal-to-noise ratio for the case where an incident, Gaussian
signal pulse made up of $q$ photons is very short compared with the
radiative lifetime and has a central frequency far removed from the
resonance frequency of the medium. If we impose the ARS criterion for 
distinguishing between superluminal propagation and propagation at the
speed of light, we find, consistent with their conclusions, that
the signal must be ``exponentially large" in order to distinguish it from
quantum noise. If the ARS criterion is replaced by a much weaker one,
however, the signal-to-noise ratio can exceed unity even for a one-photon
signal pulse, as suggested by CKK. We relate the amplified quantum
field fluctuations of ARS to quantum fluctuations of the atomic dipoles
in the case of the optical amplifier.
In Section 5, following the ideas of ARS, we present some
general considerations based on the premises of unitarity and
superluminal propagation. ARS show that, when the group velocity exceeds
the speed of light, the superluminal signal is reconstructed from a
truncated initial wavepacket, and that this truncated wavepacket has
unstable modes. We show that the truncated wavepacket introduced
by ARS propagates with both luminal and superluminal parts, and that,
while the superluminal part is the reconstructed signal, the luminal part
has the exponentially growing parts corresponding to the unstable modes.
In addition, we study the residual wavepacket formed by the difference
of the complete and truncated wavepackets. We show that contributions
from the truncated and residual wavepackets cancel in the luminal region,
but that, unlike the signal, the {\it noise} does not cancel, leading to
the conclusion that the quantum noise is mostly luminal rather than
superluminal. In the limit of a very weak incident signal pulse the
signal-to-noise ratio will be very small, consistent with the
conclusions reached by ARS.

It may be worth recalling that a primary reason for rejecting 
the possibility of superluminal transmission of information is
the requirement that causality be maintained when
Lorentz transformations are made: superluminal transmission
of information would allow an event A causing 
an event B in one reference frame to occur {\it after} event B
in a different frame. Considerations of superluminal propagation
therefore often raise questions relating to Lorentz invariance.
When and how should one include relativistic effects in order 
to ensure that physically meaningful results are obtained?

As in all previous treatments of pulse propagation in an inverted
medium that we know of, we choose the reference
frame in which the atoms are at rest. 
The Lorentz invariance of
the fundamental, fully relativistic theory implies, 
of course, that our conclusions do not depend
on this specific choice of a reference frame.
Working in this frame, 
we treat the response of the atoms to the field in the approximation
of nonrelativistic quantum mechanics. The electromagnetic field in this frame
is also treated approximately, namely in the slowly varying envelope
approximation that is used practically universally in the theory of
resonant atom-field interactions. A different choice of reference
frame would require us to start with the fully Lorentz-invariant
equations and {\it then} make the slowly-varying-envelope and 
other approximations
as appropriate. These approximations are known to be very accurate unless,
for instance, the light pulse is extremely short, and to the extent that
they are valid our results and conclusions are Lorentz-invariant. 

\section{Preliminary Considerations}
The quantum noise limitations to superluminal propagation discussed by ARS
were associated physically with spontaneous emission in the case of
an optical amplifier, and could
invalidate the CKK
results in two ways. First, CKK assume that the atoms stay in their
excited states as the pulse propagates through the amplifier. Radiative
decay of the excited state will modify their ``tachyonic dispersion relation"
and, if the decay is rapid enough, can lead to a subluminal rather
than superluminal group velocity, since $w$ in equation (\ref{eqI10}) can 
become
negative. This can be avoided by using a sufficiently short pulse.
Second, spontaneously emitted radiation
might interfere with the measurement of the superluminal group velocity
by introducing substantial noise. It is this possibility that is
addressed by ARS.

Although the ARS arguments are certainly compelling, they are based in part 
on an {\it analog} of an optical amplifier rather than a theory involving the
interaction of the electromagnetic field with an atomic medium. In
particular, theirs is a model of a single quantum field rather than coupled
atomic and electromagnetic quantum fields. The dispersion relation associated
with this model, and the criteria assumed by ARS for the observability
of superluminal propagation, 
lead to the conclusion, by
analogy to an optical amplifier, that spontaneous emission noise cannot
be avoided no matter how short the pulse or the transit time through
the amplifier. Specifically, the unstable
modes appearing in their model -- which ``are analogous to spontaneous 
emission
in the optical model of an inverted medium of two-level systems" \cite{ars} 
--
will preclude the observation of superluminal group velocity when the
pulse is made
up of a small number of photons; the quantum noise will be larger than
the signal.
In this section we present some physical considerations, motivated by the
CKK and ARS analyses, for the observability of superluminal group
velocity.

Following their equation (11), ARS
 state two {\it necessary} conditions for the observability
of superluminal propagation [$c=1$ in their units]:\\ \\
(1) $v_gT>> 1/\delta k$, where $v_g$ is the group velocity, $T$ is the
time at which the wavepacket is observed, and $\delta k$ is the spectral
width
of their initial pulse. \\ \\
(2) $(v_g-1)T>>1/\delta k$ \ .\\ \\
The first condition ensures that ``the point of observation [is] far
outside the initial spread of the wavepacket." The second allows us to
``distinguish between superluminal propagation and propagation at the
speed of light."

In the ARS model, where the field $\phi$ satisfies
\be
{\pa^2\phi\over\pa t^2}-{\pa^2\phi\over\pa z^2}-m^2\phi=0 \ ,
\label{arseq}
\ee
 the group velocity is
\be
v_g={k_o\over\sqrt{k_o^2-m^2}} \ ,
\label{p1}
\ee
where $k_o$ is the central value of the spatial frequency
$k$ for the initial pulse.
For $m<k_o$ we can approximate $v_g$ by
$1+m^2/2k_o^2$, so that condition 2 [and also condition 1] is
satisfied if
\be
m^2T>>k_o^2/\delta k>>k_o \ .
\label{p2}
\ee
$k_o>>1/T$ -- the condition that the observation time should be much larger
than the central frequency of the pulse -- then implies
\be
mT>>1 \ .
\label{p3}
\ee
Since for
$mT>>1$ the amplified quantum noise grows exponentially [see Section
3],
ARS conclude that the ``signal amplitude should be exponentially large"
in order to distinguish it from noise.
Thus, according to ARS, the observability of superluminality
 for
an input pulse consisting of only a few photons would be clouded by
spontaneous emission noise.

Consider now the implications of conditions 1 and 2 for the actual
system of interest, namely
a very short optical pulse in an inverted medium. Can we satisfy these
conditions for observation times {\it short} compared with the radiative 
lifetime?

For a short optical pulse of central frequency $\om$
propagating in an inverted medium ($w=1$)
with resonance frequency $\om_o$, the refractive index
is [equation (\ref{eqI6})]
\be
n(\om)\cong 1+{2\pi Nd^2/\hbar\over \om-\om_o}\equiv 1-{\om_p^2\over
4\om_o\Delta}
\label{p4}
\ee
for $\om_p^2/(4\om_o)<<|\om_o-\om|\equiv |\Delta|$. 
We are assuming
that $|\Delta|$ is large compared with the absorption width, which in our 
case is
the radiative decay rate.
Equation (\ref{p4}) implies
\be
{v_g\over c}=\left({d\over d\om}[\om n(\om)]\right)^{-1}=
{1\over 1-\om_p^2/4\Delta^2}
\label{p5}
\ee
and
\be
{v_g\over c}-1={\om_p^2/4\Delta^2\over 1-\om_p^2/4\Delta^2}={\om_p^2\over
4\Delta^2}{v_g\over c} \ .
\label{p6}
\ee
Then conditions 1 and 2 of ARS become, respectively,
\be
{T\over 1-\om_p^2/4\Delta^2}>>{1\over c\delta k}\sim \tau_p \ ,
\label{p7}
\ee
\be
{(\om_p^2/4\Delta^2)T\over 1-\om_p^2/4\Delta^2}>>{1\over c\delta k}\sim\tau_p
 \ ,
\label{p8}
\ee
with $\tau_p$ the pulse duration. Both conditions can be satisfied if,
for instance, 
$T>>\tau_p$ and $\om_p^2/4\Delta^2$ is not too small.
To avoid spontaneous emission during the
observation time $T$, take $T<<\tau_{RAD}$,
where $\tau_{RAD}$ is the radiative
lifetime of a single inverted atom. Then the ARS conditions require that
\be
\tau_{RAD}>>T>>\tau_p \ .
\label{p9}
\ee

As noted by CKK, there is another aspect of an inverted atomic medium that
must be addressed, namely superfluorescence (SF). 
SF is a collective phenomenon of the sample as a whole. We shall denote by
$N_T$, $S$, and $L$ the number of atoms, the cross--sectional area, and
the length of the sample, respectively, so that the density of atoms is
given by $N=N_T/ SL$.
If collisional and other dephasing mechanisms are sufficiently weak, an
inverted medium of $N_T$ atoms can emit SF radiation at the rate
\be
\tau_R=\tau_{RAD}/N_T \ ,
\ee
i.e., the radiative decay time can in effect be smaller by a factor of $N_T$
 than the
single--atom radiative lifetime $\tau_{RAD}$ assumed in the discussion thus
far.
The peak of the SF pulse occurs at a time \cite{sf}
\be
\tau_D\sim\tau_R\left[{1\over 4}\ln(2\pi N_T)\right]^2
\label{delay}
\ee
following the excitation of the atoms. It would appear then that
the quantum noise associated with SF will be small if
\be
\tau_p,L/c<\tau_R<\tau_D \ .
\label{p12}
\ee

We note for later purposes that
\be
\om_p^2={8\pi Nd^2\om_o\over\hbar}={1\over\tau_{RAD}}{N_T\over SL}Sc={
4\over\tau_R}{c\over L}
 \ ,
\label{ac}
\ee
where we have used equation (\ref{radd}) of Appendix A for the single-atom 
radiative lifetime
$\tau_{RAD}$.

This brief summary lends support to the
CKK suggestion, but obviously a more quantitative analysis
is called for. To this end we now formulate, in
the Heisenberg picture, the
quantum theory of pulse propagation in an amplifier.

\section{Formalism for
 Pulse Propagation}
We begin with the Hamiltonian for $N_T$ two--level atoms (TLAs) interacting
 with the
quantized electromagnetic field via electric dipole transitions:
\be
\hat{H}={1\over 2}\hbar\om_o\sum_{j=1}^{N_T}\sig_{zj}-d\sum_{j=1}^{N_T}\sig_
{xj}
\hat{E}(z_j)
+\sum_k\hbar\om_k\hat{a}_k^{\dag}\hat{a}_k \ ,
\label{qq1}
\ee
where $\om_o$ and $d$ have the same meaning as before and $z_j$
is the
$z$--coordinate of atom $j$. The carets (\ \^ \ ) are used to denote operators.
 We consider a one--dimensional
model
in which the atoms occupy the region from $z=0$ to $z=L$ and the field
 is a
superposition of plane waves propagating in the $z$ direction. The electric
 field
operator is given by $\hat{E}(z)=\hat{E}^{(+)}(z)+\hat{E}^{(-)}(z)$, where
\be
\hat{E}^{(+)}(z)=i\sum_k\left({2\pi\hbar\om_k\over S\ell}\right)^{1/2}
\hat{a}_ke^{i
kz} \ \
\ \ (k=\om_k/c)
\label{qq2}
\ee
and $\hat{E}^{(-)}(z)=\hat{E}^{(+)}(z)^{\dag}$. $S\ell$, where $S$,
as before,
is a cross--sectional area and
$\ell$ a length, is the quantization volume. For simplicity we consider only
 a single
field polarization, namely linear polarization along the direction of the
 transition
dipole moment of the TLAs. $\hat{a}_k$ and $\hat{a}_k^{\dag}$ are the photon
annihilation and creation operators, respectively, for mode $k$, and the
$\sig$'s are
the Pauli two--state operators in the standard notation.

We will work in the Heisenberg picture, in which the time--dependent electric
 field
operator satisfies
\be
\left({\pa^2\over\pa z^2}-{1\over c^2}{\pa^2\over\pa t^2}\right)\hat{E}={4\pi
\over
c^2}{\pa^2\hat{P}\over\pa t^2}={4\pi d\over c^2S}\sum_{j=1}^{N_T}{\pa^2\sig_{xj}
\over\pa t^2}
\delta(z-z_j)
\rightarrow {4\pi\over c^2}Nd{\pa^2\over\pa t^2}\sig_x(z,t) \ ,
\label{maxwell}
\ee
where in the last step we have made the continuum approximation for the
 polarization
density $\hat{P}$, assuming a uniform atomic density $N$. We now write
\be
\hat{E}^{(+)}(z,t)=\hat{F}(z,t)e^{-i\om(t-z/c)}
\label{Fdef}
\ee
and assume $\hat{F}(z,t)$ is slowly varying in $z$ and $t$ compared with
$\exp[-i\om(t-z/c)]$. In this
approximation
\be
2i{\om\over c}\left({\pa \hat{F}\over\pa z}+{1\over c}{\pa \hat{F}
\over\pa t}\right)
+ h.c.={4\pi\over
c^2}Nd{\pa^2\sig_x\over\pa t^2}e^{i\om(t-z/c)} \ .
\label{q1}
\ee
It will be convenient to use the atomic lowering and raising
operators
$\sig={1\over 2}(\sig_x-i\sig_y)$ and
$\sig^{\dag}={1\over 2}(\sig_x+i\sig_y)$, respectively, such that $[\sig,
\sig^{\dag}]
=-\sig_z$, and
to write
\be
\sig(z,t)=\hat{s}(z,t)e^{-i\om(t-z/c)} \ ,
\label{sdef}
\ee
where the operator
$\hat{s}(z,t)$ is assumed to be slowly varying in the same 
sense as $\hat{F}(z,t)$. 
Then,
 in
the rotating--wave approximation (RWA), we can replace (\ref{q1}) with
\be
{\pa \hat{F}\over\pa z}+{1\over c}{\pa \hat{F}\over\pa t}=
(2\pi iNd{\om_o\over c})\hat{s} \
 ,
\label{q2}
\ee
where on the right-hand side we have approximated $\om$ by $\om_o$.
This equation and the TLA Heisenberg equations
\be
{\pa \hat{s}\over\pa t}=-i(\Delta-i\beta)\hat{s}-{id\over\hbar}\sig_z\hat{F} \ ,
\label{q4}
\ee
\be
{\pa {\sig}_z\over\pa t}=-2\beta(1+\sig_z)-{2id\over\hbar}(\hat{F}^{\dag}
\hat{s}-
\hat{s}^{\dag}\hat{F}) \ ,
\label{q5}
\ee
derived in Appendix A form a closed set of operator equations. 
They provide the basis
for a quantum theory of propagation in either amplifying or absorbing media.

In the semiclasical approximation in which the atom and field operators are
replaced by their expectation values, equations (\ref{q2})-(\ref{q5}) reduce
to well known Maxwell-Bloch equations. 
Otherwise, different limits can apply:
\begin{itemize}
\item The limit of $\beta\rightarrow 0, \Delta=0$ and
$\sig_z\rightarrow 1$ considered below gives equations
(\ref{q6})-(\ref{q8}) implying superfluorescence when the initial state
of the field is the vacuum.
\item The limit of $\omega\gg\omega_0$ gives the ARS
field equation, as discussed below.
\item Finally, in Section 4 the CKK case of large detuning,
$\sig_z\rightarrow 1$, and the initial state of a very short incoming
pulse, is studied.
\end{itemize}

If the field central frequency $\om$ is assumed to match exactly the atomic
resonance frequency $\om_o$, so that $\Delta=0$, and if we restrict 
ourselves
to times short compared with the single--atom radiative
lifetime $[\tau_{RAD}=(2\beta)^{-1}]$
and assume that the atoms remain with probability $\cong 1$
 in their excited states over times of interest, we can ignore
(\ref{q5}) and replace $\sig_z(z,t)$
by 1 and equation (\ref{q4}) by
\be
{\pa \hat{s}\over\pa t}=-{id\over\hbar}\hat{F} \ .
\label{q5a}
\ee
In terms of the independent variables $\zeta=t-z/c$ and $\eta=z$,
\be
{\pa \hat{s}\over\pa\zeta}=-{id\over\hbar}\hat{F}  \ ,
\label{q6}
\ee
\be
{\pa \hat{F}\over\pa\eta}=\left(2\pi iNd{\om_o\over c}\right)\hat{s} \ ,
\label{q7}
\ee
implying
\be
{\pa^2\hat{s}\over\pa\eta\pa\zeta}=\left({\om_p^2\over 4c}\right)\hat{s}
 \ , \ \ \
 {\pa^2\hat{F}\over\pa\eta\pa\zeta}=\left({\om_p^2\over 4c}\right)\hat{F} \ .
\label{q8}
\ee
Equations (\ref{q6})-(\ref{q8}) have been used in studies of the
buildup of superfluorescent radiation \cite{sf}. It will be useful for the
discussion in Section 4
 to briefly rederive here one of the most important
results of those studies.

Equation (\ref{q2}) has the formal solution
\bea
\hat{F}(z,t)&=&\hat{F}_o(z,t)+\left(2\pi iNd{\om_o\over c}\right)
\int_0^zdz'\hat{s}(z',
t-{z-z'\over
c})\theta(t-{z-z'\over c})  \nonumber \\
&=&\hat{F}_o(z,t)+\left(2\pi iNd{\om_o\over
c}\right)\int_0^zdz'\hat{s}(z-z',t-z'/c)\theta(t-z'/c)  \ ,
\label{q9}
\eea
where we have chosen the retarded Green function over the advanced
 Green function in order to ensure causality.
Here $\theta$ is the unit step function and $\hat{F}_o(z,t)$ is a solution
of the homogeneous equation. We are interested here in the
expectation value $\la \hat{F}^{\dag}(L,t)\hat{F}(L,t)\ra$ at the end ($z=L$)
of the medium. For SF the expectation value is taken over the vacuum state
of the field, in which case the first term on the right-hand side of 
(\ref{q9})
does not contribute to normally ordered expectation values. We
may therefore ignore this term for practical purposes. Defining
$y=2\sqrt{\zeta\eta}$ we
 find
from (\ref{q8}) that $\hat{s}$ satisfies the differential
equation
for $I_0(y)$, the modified Bessel function of order zero \cite{bc}. The 
solution
of interest for $\hat{F}(L,t)$ is then \cite{sf}
\be
\hat{F}(L,t)=\left(2\pi iNd{\om_o\over
c}\right)\int_0^Ldz'\hat{s}(L-z',0)I_0\left(\om_p\sqrt{(z'/c)(t-z'/c)}\right)
\theta(t-z'/c) \ .
\label{q10}
\ee

In order to calculate $\la \hat{F}^{\dag}(L,t)\hat{F}(L,t)\ra$ we require
$\la \hat{s}^{\dag}(z',0)
\hat{s}(z,0)\ra$, which we evaluate in Appendix B.
We obtain \cite{sf}
\be
\la \hat{F}^{\dag}(L,t)\hat{F}(L,t)\ra=\left(2\pi d{\om_o\over c}\right)^2
{N\over S}
\int_0^Ldx\theta(t-x/c)I_0^2\left(\om_p\sqrt{(x/c)(t-x/c)}\right) \ .
\label{q11}
\ee
For times large enough that $I_0$ may be replaced by its
asymptotic form,
\be
\la \hat{F}^{\dag}(L,t)\hat{F}(L,t)\ra\sim {1\over 8\pi}{2\pi\hbar\om_o
\over Sct}
e^{\sqrt{t/\tau_R}} \ .
\label{q111}
\ee
Equating the intensity expectation value $(c/2\pi)\la \hat{F}^{\dag}(L,t)
\hat{F}(L,t)\ra$
to the maximum expected SF intensity $N_T\hbar\om_o/S\tau_R$, we arrive at
the expression (\ref{delay}) for the time at which the SF pulse reaches its peak
intensity. In the short-time limit, on the other hand,
\be
\la \hat{F}^{\dag}(L,t)\hat{F}(L,t)\ra\sim \left(2\pi d{\om_o\over c}
\right)^2{N\over S}
ct \ ,
\label{stl}
\ee
a result we will return to in Section 4.

\subsection*{Approximation Leading to ARS Field Equation}
Our considerations thus far assume that the field
central frequency lies in the vicinity of the atomic resonance in the sense 
that the
detuning $\Delta$ is small in magnitude compared with $\om$ and $\om_o$. Let
 us now
suppose instead that the field frequency $\om$ is very large compared
with
$\om_o$. In this case we must work with the atomic
operators $\sig_x$, $\sig_y$ instead of the slowly varying $\hat{s}$.
>From equations (\ref{sigx1}) and (\ref{sigy1})
of Appendix A we have
\be
\ddot{\sig}_x+\om_o^2\sig_x=-{2d\om_o\over\hbar}\sig_z\hat{E}\cong
-{2d\om_o\over\hbar}\hat{E}
\ee
in the approximation $\sig_z\cong 1$.
The assumption
$\om>>\om_o$ implies
\be
\ddot{\sig}_x\cong -{2d\om_o\over\hbar}\hat{E} \ ,
\ee
so that, from equation (\ref{maxwell}),
\be
\left({\pa^2\over\pa t^2}-c^2{\pa^2\over\pa z^2}-\om_p^2\right)\hat{E}=0 \ .
\label{arsapp}
\ee
This is identical to the equation of motion for the quantum field in the ARS
 model
when we equate $\om_p^2$ to their $m^2$. From this perspective the
 ARS
equation of motion describes the interaction of the electromagnetic field 
with $N$
unbound electrons ($\om>>\om_o$)
per unit volume. However, the usual plasma
dispersion formula $n^2=1-\om_p^2/\om^2$ for the refractive index $n$
 is replaced in this case by
\be
n^2=1+\om_p^2/\om^2 \ .
\ee
This is a consequence of the assumption $\sig_z\cong 1$; had we assumed 
$\sig_z
\cong
-1$ we would have obtained the familiar plasma dispersion formula.

To describe the growth of
the quantum noise with time in this model, we write (\ref{arsapp})
in the form
\be
{\pa^2\hat{E}\over\pa\tau_1\pa\tau_2}-{m^2\over 4}\hat{E}=0 \ ,
\label{treq}
\ee
where $\tau_1=t-z/c$, $\tau_2=t+z/c$. In terms of the independent variable
$y=m\sqrt{\tau_1\tau_2}$, equation(\ref{treq}) has solutions that are 
linear
combinations of the zero-order modified Bessel functions $I_0(y),K_0(y)$.
For large $t$, the vacuum expectation value
\be
\la \hat{E}^2(z,t)\ra\propto I_0^2(y)\sim{e^{2mt}\over 2\pi mt} \ ,
\label{arsnoise}
\ee
so that the quantum noise grows exponentially in time from the initial
fluctuations of the vacuum field, the fluctuations present before
the medium in the ARS model is ``inverted."

\section{Signal and Noise}
We wish to determine to what extent the observation of the superluminal
group velocity considered by CKK will be affected by quantum noise. The 
system
of interest is described by the Heisenberg equations of motion (\ref{q2}) and
(\ref{q4}). We approximate $\sig_z$ by 1, assuming that
pulse durations $\tau_p$ and transit times $L/c$ are sufficiently small
that de-excitation of the initially inverted atoms by radiation (or any other
decay process) is negligible. The situation here is different from that
describing the onset of SF in that (a) the detuning $\Delta$ is not zero
but is instead large (Section 2), and (b) the initial state of the field
is not the vacuum but corresponds to a short pulse of radiation from some
external source.

The equation for $\hat{s}(z,t)$ in the present model is
\be
{\pa \hat{s}\over\pa t}=-i(\Delta-i\beta)\hat{s}-{id\over\hbar}\hat{F} \ ,
\label{qqq1}
\ee
or
\be
\hat{s}(z,t)=\hat{s}(z,t_o)e^{-i(\Delta-i\beta)(t-t_o)}-{id\over\hbar}
\int_{t_o}^tdt'\hat{F}
(z,t')
e^{i(\Delta-i\beta)(t'-t)} \ .
\label{qqq2}
\ee
$t_o$ is some initial time, before any pulse is injected
into the medium. We take $\hat{F}(z,t_o)=0$, although of course
what this really means is that there is no nonvanishing field or intensity
in the medium at $t_o$, so that for practical purposes 
(normally ordered expectation values) we can in effect
ignore the operator $\hat{F}(z,t_o)$ in the equation for $\hat{s}(z,t)$.

The pulse is assumed to have a central frequency $\om$ and to have no
significant frequency components near the resonance frequency $\om_o$:
 $|\Delta|\tau_p>1$. We assume that $|\Delta|\tau_p$ is large
enough that
we can approximate (\ref{qqq2}) by integrating by parts and retaining only 
the leading terms:
\be
\hat{s}(z,t)\cong \hat{s}(z,t_o)e^{-i(\Delta -i\beta)(t-t_o)}-
{d\over\hbar}{\Delta+i\beta
\over
\Delta^2+\beta^2}\hat{F}(z,t)-{id\over\hbar\Delta^2}{\pa \hat{F}\over\pa t} \ .
\label{qq3}
\ee
As will be clear from the analysis that follows, this approximation
implies the undistorted propagation of the incident pulse at the group
velocity $v_g$, as assumed by CKK.

>From (\ref{q2}),
\be
{\pa \hat{F}\over\pa z}+{1\over c}{\pa \hat{F}\over\pa t}\cong 
\left(2\pi iNd{\om_o\over
c}\right)\hat{s}(z,t_o)e^{-i(\Delta-i\beta)(t-t_o)}+{g\over 2}\hat{F}
+i[n(\om)-1]{\om
\over c}\hat{F}
+\left({1\over c}-{1\over v_g}\right){\pa \hat{F}\over \pa t} \ ,
\label{qq4}
\ee
where
\be
g\equiv {4\pi Nd^2\om_o\over\hbar c}{\beta\over\Delta^2+\beta^2}
\label{qq5}
\ee
is the gain coefficient for propagation of a field with frequency
$\om$ in the inverted medium. We have used equation (\ref{p4}) for the
refractive index $n(\om)$ and (\ref{p6}) for $v_g/c-1$.
Writing
$\hat{F}(z,t)=\hat{F}'(z,t)e^{i[n(\om)-1]\om z/c}$ and $\hat{s}(z,t_o)=
\hat{s}'(z,t_o)e^{i[n(\om
)-1]\om z/c}$
yields an equation in terms of the primed variables in which the term
$i[n(\om)-1](\om/c)z$ associated with phase velocity is eliminated. Then,
ignoring for practical purposes the difference between the primed and
unprimed variables, we have
\be
{\pa \hat{F}\over\pa z}+{1\over v_g}{\pa \hat{F}\over\pa t}= 
{g\over 2}\hat{F} +
\left(2\pi iNd{\om_o\over
c}\right)\hat{s}(z,t_o)e^{-i(\Delta-i\beta)(t-t_o)} \ ,
\label{qq6}
\ee
and therefore
\bea
\hat{F}(z,t)&=&\hat{F}(0,t-z/v_g)e^{gz/2}+
\left(2\pi iNd{\om_o\over c}\right)\int_0^zdz'
\hat{s}(z',t_o)e^{g(z-z')/2} \nonumber \\
&\times& e^{-i(\Delta-i\beta)[t-t_o-(z-z')/v_g]}\theta(t-t_o-(z-z')/v_g) 
\nonumber \\
&\equiv& \hat{F}_s(0,t-z/v_g)e^{gz/2}+\hat{F}_n(z,t) \ ,
\label{qq7}
\eea
where the subscripts s and n denoted signal and noise, respectively. Here
\be
\hat{F}_n(z,t)=\left(2\pi iNd{\om_o\over c}\right)\int_0^zdz'
\hat{s}(z',t_o)e^{g(z-z')/2}e^{-i(\Delta-i\beta)[t-t_o-(z-z')/v_g]}
\theta(t-t_o-(z
-z')/v_g)
\label{qq8}
\ee
is a quantum noise field associated with the quantum fluctuations of the 
atomic dipoles.

To appreciate the significance of $g$ as defined by equation (\ref{qq5}),
consider the
gain coefficient $g_R$ for a radiatively broadened transition of frequency $
\om_o$ and
radiative decay rate $1/\tau_{RAD}=2\beta$. For light of frequency $\om=
\om_o-\Delta$,
\be
g_R={NS\over\tau_{RAD}}{2\beta\over\Delta^2+\beta^2}
={4\pi
Nd^2\om_o\over\hbar c}{\beta\over\Delta^2+\beta^2}
\ee
if we assume that all the $N$ atoms per unit volume are in the upper state of
 the
amplifying transition. Thus $g_R=g$, i.e., $g
$ is just the
gain coefficient for amplification by stimulated emission. We note also that,
from equation (\ref{p6}),
\be
g=2\beta\left({1\over c}-{1\over v_g}\right)
\label{qq9}
\ee
in the case under consideration where the amplifying transition is
 radiatively
broadened and the detuning is large compared with the gain bandwidth.

The operator $\hat{s}(z,t_o)$ has the expectation-value properties 
 (\ref{cor1}) and (\ref{cor2})
of Appendix B.
These properties imply $\la \hat{F}_n(z,t)\ra=
\la \hat{F}_n^{\dag}(z,t)\ra=0$ and
\bea
\la \hat{F}_n^{\dag}(z,t)\hat{F}_n(z,t)\ra&=&\left(2\pi 
Nd{\om_o\over c}\right)^2{L
\over N_T}
e^{-2\beta(t-t_o)}\int_{z-v_g(t-t_o)}^zdz'e^{g(z-z')}e^{2\beta(z-z')/v_g}
\nonumber \\
&=&\left(2\pi d{\om_o\over c}\right)^2{N\over
S}{c\over 2\beta}\left[e^{gv_gt}-e^{-2\beta t}\right] \ ,
\label{qq12}
\eea
where we have used the relations (\ref{qq9}) and $N_T=NSL$ and, to
simplify the notation,
we have taken $t_o=0$.

Since the atom and field are initially uncorrelated, i.e.,
\be
\la \hat{F}^{\dag}(0,t-z/v_g)\hat{s}_j(t_o)\ra=\la \hat{F}_n^{\dag}
(0,t-z/v_g)\ra\la \hat{s}_j(t_o)
\ra=0 \ ,
\label{qa1}
\ee
we have, at the end of the amplifier,
\be
\la \hat{F}^{\dag}(L,t)\hat{F}(L,t)\ra=\la \hat{F}^{\dag}_s(0,t-L/v_g)
\hat{F}_s(0,t-L/v_g)\ra e^{gL}
+\la \hat{F}^{\dag}_n(L,t)\hat{F}_n(L,t)\ra
\label{qa2}
\ee
and the signal--to--noise ratio
\bea
SNR(L,t)&\equiv&{\la \hat{F}^{\dag}_s(0,t-L/v_g)\hat{F}_s(0,t-L/v_g)\ra 
e^{gL}\over\la
\hat{F}^{\dag}_n(L,t)\hat{F}_n(L,t)\ra} \nonumber \\
&=&{\la \hat{F}_s^{\dag}(0,t-L/v_g)\hat{F}_s(0,t-L/v_g)\ra e^{gL}\over
(2\pi d\om_o/c)^2(N/S)(c/2\beta)\left[e^{gL}-e^{-2\beta L/v_g}\right]}
\nonumber \\
&=&{\la \hat{F}^{\dag}_s(0,t-L/v_g)\hat{F}_s(0,t-L/v_g)\ra 
e^{2\beta(1/c-1/v_g)L}\over
(2\pi
d\om_o/c)^2(N/S)(c/2\beta)\left[e^{2\beta(1/c-1/v_g)L}-
e^{-2\beta L/v_g}\right]}
\nonumber \\
&\cong&{\la \hat{F}^{\dag}_s(0,t-L/v_g)\hat{F}_s(0,t-L/v_g)
\ra\over(2\pi d\om_o/c)^2NL/S
} \ .
\label{qa3}
\eea
In the denominators
we have taken $t=L/v_g$ for the time over which the atoms radiate,
and have used the fact that $2\beta(1/c-1/v_g)L=gL$, the
difference of two numbers that themselves are small according to our 
assumption that
propagation times are small compared with the single--atom radiative decay 
rate, is $<<1$.

The numerator in equation (\ref{qa3}) can be related to the expectation value
 $q$ of the
number of photons in the incident signal pulse as follows.
The expectation
value of the incident signal intensity is
\be
I_s(0,t)={v_g\over 2\pi}\la \hat{F}_s(0,t)\hat{F}_s(0,t)\ra=
I_oe^{-t^2/\tau_p^2}
\ee
for a Gaussian pulse of duration $\tau_p$. Requiring that the energy
flux
$\int_{-\infty}^{\infty}dtI_s(z,t)$ be $q\hbar\om/S\cong q\hbar\om_o/S$
implies $I_o=q\hbar\om_o/(S\tau_p\sqrt{\pi})$ and therefore
\be
\la \hat{F}^{\dag}_s(0,t-L/v_g)\hat{F}_s(0,t-L/v_g)\ra=q{2\pi\hbar\om_o\over
v_gS\tau_p\sqrt{\pi}}e^{-(t-L/v_g)^2/\tau_p^2} \ .
\ee
Thus
\bea
SNR(L,t)&=&{q\over\tau_p\sqrt{\pi}}{c\over v_g}e^{-(t-L/v_g)^2/\tau_p^2}
\left({2\pi
d^2\om_o\over\hbar cS}NSL\right)^{-1} \nonumber \\
&=&{q\over\sqrt{\pi}}\left({4c\over\om_p^2L\tau_p}\right){c\over v_g}
e^{-(t-L/v_g)^2/\tau_p^2} \nonumber \\
&=&{q\over\sqrt{\pi}}{\tau_R\over\tau_p}{c\over v_g}
e^{-(t-L/v_g)^2/\tau_p^2} \ ,
\label{qa4}
\eea
where we have used equation (\ref{ac}) \cite{pi}.

Among the criteria given by CKK for the observation of a superluminal
pulse is that ``The probe-pulse duration [$\tau_p$] must not exceed
$\tau_R=4c/L\om_p^2$." This criterion implies, from equation
(\ref{qa4}), that 
$SNR(L,t)\geq(q/\sqrt{\pi})c/v_g$ and therefore that it is possible,
even for $q\sim 1$, to have superluminal propagation with
$SNR(L,t)>1$ if the pulse duration is short enough: $\tau_p<\tau_R
c/v_g$.

In order to relate this conclusion to ARS,
we use equation (\ref{p6}) to write (\ref{qa4}) as
\be
SNR(L,t)={q\over\sqrt{\pi}}{\tau_p\over \left({v_g\over c}-1\right)
{L\over c}\Delta^2\tau_p^2}e^{-(t-L/v_g)^2/\tau_p^2} \ .
\label{new1}
\ee
We see from this expression that, if we impose the ARS condition (2),
i.e., $(v_g/c-1)L/c>>\tau_p$, then
\be
SNR(L,t)<<{q\over\sqrt{\pi}}{1\over(\Delta\tau_p)^2}e^{-(t-L/v_g)^2/
\tau_p^2} \ ,
\ee
so that, given also the condition on $|\Delta|\tau_p$ discussed before
equation (\ref{qq3}), {\it the signal-to-noise ratio will be very small
when the ARS condition for strong distinguishability of superluminal
propagation from propagation at the speed $c$ is satisfied.}

In fact if $(v_g/c-1)L/c>>\tau_p$ and therefore $SNR(L,t)$ is very small
for $q\approx 1$, then
\be
t/\tau_R={L/c\over\tau_R}={v_g\over c}{L/v_g\over\tau_R}\stackrel{>}{\sim}
{v_g\over c}{\tau_p\over\tau_R} \ ,
\label{new2}
\ee
which, from (\ref{qa4}), must be large. Then the SF noise
must be exponentially large [equation (\ref{q111})]. It follows 
that $q$ must be exponentially 
large in order to maintain a signal-to-noise ratio greater than
unity. This is consistent with the ARS conclusion that ``for the signal
amplitude to be larger than the amplitude of the fluctuations at the
observation time, the signal amplitude should be exponentially large"
\cite{ars}. 

Our results are therefore in agreement with ARS in that, if we require
the separation of the superluminal pulse and a twin vacuum-propagated
pulse to be much larger than the pulse duration, the signal-to-noise
ratio will be very small at the one- or few-photon level. On the
other hand, the results are not inconsistent with CKK:
 even at the one-photon level we can achieve a signal-to-noise
ratio greater than unity if this separation ($[v_g/c-1]L/c$) is smaller than the
pulse duration $\tau_p$ [Equation (\ref{new1})].

\subsection*{Physical Origin of the Noise Limiting the Observation of \newline
Superluminal Group Velocity} 

Note that, when we set the
time $t$ in equation (\ref{stl}) for the short-time
SF noise intensity equal to
the ``observation time" $L/c$, we obtain exactly the noise
intensity appearing in the denominator of equation (\ref{qa3})
\cite{remark2}. Thus the quantum noise
that imposes limitations on the observation of superluminal group
velocity is attributable to the initiation of SF.

\subsection*{Operator Ordering and Relation to ARS Approach} 
Less obvious, perhaps, is the relation between the quantum noise
we have considered -- which stems from the {\it atomic dipole
fluctuations} characterized by equations (\ref{cor1}) and (\ref{cor2}) 
of Appendix B -- and the quantum noise of ARS, which is attributed
to the {\it quantum fluctuations of the field}.

To establish the relation to the ARS approach we return to
our calculation of the noise intensity, using now {\it anti}-normally
ordered field operators instead of the normally ordered operators
used before. Thus we consider now the expectation value
$\la \hat{F}(z,t)\hat{F}^{\dag}(z,t)\ra$ instead of 
$\la \hat{F}^{\dag}(z,t)\hat{F}(z,t)\ra$.
In this approach the atomic dipole fluctuations play no explicit
role, as can be seen from equation (\ref{qq7}) and the fact that 
\be
\la \hat{s}(z',t_o)\hat{s}^{\dag}(z'',t_o)\ra=0
\label{noise1}
\ee
for excited atoms. In this case, however, the initially unoccupied
modes of the field make a nonvanishing contribution as a
consequence of non-normal ordering:
\bea
\la \hat{F}(0,t-L/v_g)\hat{F}^{\dag}(0,t-L/v_g)\ra&=
&\sum_k{2\pi\hbar\om_k\over S\ell}
\la \hat{a}_k(0)\hat{a}_k^{\dag}(0)\ra e^{g(\om_k)L} \nonumber \\
&\cong&\sum_k{2\pi\hbar\om_k\over S\ell}[g(\om_k)L+1] \ ,
\label{noise2}
\eea
which follows from (\ref{qq2}) and (\ref{Fdef}) and the approximation 
$gL<<1$ upon which (\ref{qa4}) is based. The contribution from the
term that does not vanish as $L\rightarrow 0$ can be ignored, as it
corresponds to vacuum quantum noise (energy ${1\over 2}\hbar\om_k$ per mode)
that is present even in the absence of the amplifier. In other words, the
quantum noise of the field in the presence of the amplifier is
\bea
\la \hat{F}(0,t-z/v_g)\hat{F}^{\dag}(0,t-z/v_g)\ra_n
\equiv\sum_k{2\pi\hbar\om_k\over
S\ell}g(\om_k)L 
\rightarrow{\ell\over 2\pi c}\int d\om{2\pi\hbar\om\over S\ell}g(\om)L 
\nonumber \\
\cong\pi\left({2\om_od\over c}\right)^2{NL\over c}
\int_0^{\infty}d\om
{\beta\over\Delta^2+\beta^2} \ , 
\label{noise3}
\eea
where we have gone to the mode
continuum limit, approximated $\om$ by $\om_o$ in the numerator of the
integrand, and used equation (\ref{qq5}) for the gain coefficient.
Performing the integration, we obtain exactly the noise term appearing in
the denominator in the last line of (\ref{qa3}).  
{\it But now the noise is attributable to the
amplification of vacuum field fluctuations} \cite{remark}.

Thus we can attribute the quantum noise that limits the observation of
superluminal group velocity to either the quantum fluctuations of
the field in the inverted medium, as do ARS, or to the quantum
fluctuations of the inverted atoms, as in our derivation
of the signal-to-noise ratio. The situation here is similar to
that in the theory of the initiation of SF, as discussed by Polder
et al. \cite{sf}, or, as noted by those authors, to the theory of
spontaneous emission by a single atom \cite{mas}.

\subsection*{Limit of Very Small Transition Frequency}
Since the origin of noise in the optical amplifier is associated
ultimately with spontaneous emission, the question arises as to
whether the signal-to-noise ratio might be increased by employing a
transition having a very small transition frequency $\om_o$ and
therefore a very large radiative lifetime. Indeed, since $\om_p^2\propto
\om_o$, the second line of equation (\ref{qa4}) suggests at first
glance that $SNR\rightarrow\infty$ in the limit $\om_o\rightarrow 0$. However,
equation (\ref{p5}) shows that $v_g\rightarrow c$ in this limit: 
the superluminal effect itself becomes weaker as 
the spontaneous emission rate is made smaller. 

In this connection we invoke once again the form (\ref{new1}) of
the signal-to-noise ratio. If we assume $|\Delta|\tau_p>1$ in order that
the pulse does not undergo substantial distortion as a consequence
of strong absorption, then
\be
SNR(L,t)<q{c\tau_p\over (v_g-c)L/c} \ .
\ee
In other words, the signal-to-noise ratio must be smaller than the
number of photons in the incident pulse times a factor equal to
to length of the vacuum-propagated pulse divided by the separation of
the vacuum-propagated pulse and the pulse emerging from the amplifier,
{\it independent of the the atomic transition frequency or the radiative
lifetime}. At the one- or few-photon level the signal-to-noise ratio
must therefore be less than unity under the ARS criteria for the
observation of superluminal group velocity, regardless of the frequency
or strength of the amplifying transition. 

\section{Unitarity and Superluminal Propagation}
We now turn our attention from the specific example
of the optical amplifier to some general features of superluminal
propagation that follow generally from the unitary evolution of
the state vector, considered here within first quantization. 

The time evolution  of a
wave packet can be formulated in terms of a
unitary operator
$U(t)$ or equivalently in terms of a
coordinate-space propagator $G(x-x',t)=\la x
|U(t)| x'\ra$:
\bea
|\Psi(t)\ra&=&U(t)|\Psi(0)\ra \label{U} \nonumber \\
\Psi(x,t) &=&\la x|\Psi(t)\ra=\int^{\infty}_{-\infty} d x' G(x-x',t) \Psi(x',0) .
\label{eqII1}
\eea
The assumption that
the propagator vanishes identically outside the
light cone implies that
\bea
G(x-x'>ct,t)=0 .
\label{eqII2}
\eea
Given an initial wave packet centered around
$x=X_0<0$ at $t=0$, we assume that at a later time
$t>0$ it will be centered  around
$X_0+v_g t$, as in the example of pulse propagation
in an inverted medium.

We divide the wave packet into two parts, which we label as
``superluminal" (S) and
``luminal" (L), in the following way:
\bea
\Psi(x,T)\equiv \left\{ \begin{array}{ll}
\Psi^S(x,T) & \ \ x>cT \ , \\
\Psi^L(x,T) & \ \ x<cT  \ . \label{eqII3}
\end{array} \right.
\eea
$\Psi^S$ vanishes if the group velocity $v_g<c$.

Suppose that $v_g>c$ and that we let the
wave packet propagate for a time $T$ long enough
that a superluminal signal can be clearly
identified.
That is, we assume that at $t=T$,
\be
\Psi^L(x,T)\approx 0. \label{eqII4}
\ee
Now
\be
\la \Psi(T)|\Psi(T)\ra = \la \Psi(0)|\Psi(0)\ra
\ee
due to unitarity, and thus
\be
\int^{\infty}_{-\infty} d x' |\Psi(x',0)|^2
=\int^{\infty}_{-\infty} d x' |\Psi(x',T)|^2
\approx\int^{\infty}_{cT} d x' |\Psi^S(x',T)|^2 \ .
\label{eqII4a}
\ee
Physically, this means that the
superluminal signal, $\Psi^S(x,T)$, is about as large,
or  contains about ``as many photons," as the initial wave packet.

We now combine the two underlying premises of causality and 
superluminal propagation as they are defined by equations (\ref{eqII2})
and (\ref{eqII4}). Using equation (\ref{eqII1}) for $x>cT$, we write
\be
\Psi^S(x,T)=
\int^{0}_{-\infty} d x' G(x-x',T) \Psi(x',0)
+ \int^{\infty}_{0} d x' G(x-x',T) \Psi(x',0) .
\ee
The first term vanishes because, according to
Eq. (\ref{eqII2}), the  integrand differs from
zero only if $x'>x-cT>0$. Thus
\be
\Psi^S(x,T)=
\int^{\infty}_{-\infty} d x' G(x-x',T)
\left[ \Theta(x') \Psi(x',0) \right] . \label{eqII5}
\ee
This formulates the notion, which is essential to the ARS argument, that 
 for a {\it
causal}, [i.e., equation (\ref{eqII2})],
superluminal signal [equation (\ref{eqII4})], {\it the
wave packet is reconstructed from its tail}
[equation (\ref{eqII5})]. 

This rather remarkable reconstruction of the signal propagated without
distortion and with {\it superluminal} group velocity is especially
evident in the temporal domain \cite{diener}. (See Figure 1.)

The construction (\ref{eqII5}) of the
superluminal wave packet
from the tail of the initial wave packet motivated ARS to define
another, {\it truncated} initial wave packet:
\be
\Phi(x,0)\equiv \Theta(x) \Psi(x,0) .
\ee
The two {\it different} initial wave functions, $\Phi(x,0)$ and
$\Psi(x,0)$, give the same superluminal signal:
\be
\Psi(x>cT,T)= \Phi(x>cT,T)=
\int^{\infty}_{-\infty} d x' G(x-x',T)
\Phi(x',0) . \label{eqII6}
\ee

Equation (\ref{eqII6}) implies what ARS
call {\it amplification}: a ``small" signal propagates to become a
``large'' signal. After all, $\Phi(x,0)$ is ``made from a small number of
photons,'' while we have just seen that $\Psi(x>cT,T)$ has about
 the same number
of photons as the non-truncated initial wave packet. We note that
{\it amplification in this sense is a necessary consequence of a superluminal 
group velocity}.

One might be tempted to write
(\ref{eqII6}) symbolically as
\be
|\Phi(0)\ra \longrightarrow |\Psi^S(T)\ra ,
\label{wrong1}
\ee
where $\longrightarrow$ denotes time
evolution under $U(T)$.
This would be incorrect:
the truncated initial wave packet $\Phi(x,0)$ is a
perfectly well defined initial state, but it {\it
does not} evolve into $\Psi^S(x,T)$; part of it evolves
luminally. It will be prove convenient
to introduce ``superluminal" and ``luminal" parts of the
truncated wave packet in a manner similar to the decomposition (\ref{eqII3})
used for the complete wave packet
$\Psi(x,T)$:
\bea
\Phi(x,T)\equiv \left\{ \begin{array}{ll}
\Phi^S(x,T) & \ \ x>cT \ , \\
\Phi^L(x,T) & \ \ x<cT  \ .
\end{array} \right.
\label{eqII7}
\eea
We note that, while the superluminal part of
the time-evolved  truncated initial state is the
same as  the superluminal part of the
time-evolved non-truncated initial state, the
luminal parts of these signals differ:
\bea
&&\Phi^S(x,T)=\Psi^S(x,T), \\
&&\Phi^L(x,T)\neq\Psi^L(x,T)\approx 0.
\label{eqII8}
\eea
That is, while the luminal part of the time-evolved
complete wave packet approximately vanishes
($\Psi^L(x,T)\approx 0$), the luminal part of the
truncated wavepacket, $\Phi^L(x,T)$, does
not. We show below that, on the contrary, it
grows exponentially with time.

\subsection*{Momentum space: normal and unstable
modes}

We are comparing the time evolution of two
different initial wavepackets, $\Psi(x,0)$ and
$\Phi(x,0)$ where
$\Phi(x,0)=\Theta(x)\Psi(x,0)$.
It is useful to define still another initial wave packet,
\be
R(x,0)=\Theta(-x)\Psi(x,0) \ .
\label{IIa1}
\ee
Clearly,
\be
\Psi(x,0)=R(x,0)+\Phi(x,0) \ .
\label{eqIIa2}
\ee
After a time $T$,
$\Psi(x,0)$ evolves into $\Psi(x,T)$,
$\Phi(x,0)$ into $\Phi(x,T)$, and
$R(x,0)$ into $R(x,T)$.
The time evolution is linear and
\be
\Psi(x,T)=R(x,T)+\Phi(x,T) \ .
\label{eqIIa3}
\ee

Fourier transforming into momentum space,
we define
$g(k)$,
$\zeta(k)$, and $\xi(k)$, by:
\bea
&&\Psi(x,t)\equiv\int_{-\infty}^{\infty} dk g(k)
\exp[i (kx-\omega_k t)] ,
\label{g} \\
&&\Phi(x,t)\equiv\int_{-\infty}^{\infty} dk \zeta(k)
\exp[i(kx-\omega_k t)],
\label{zeta} \\
&&R(x,t)\equiv\int_{-\infty}^{\infty} dk \xi(k)
\exp[i(kx-\omega_k t)] \label{xi} .
\eea
>From these definitions it is straightforward
to show that
\bea
&&\zeta(k) =\frac{-i}{2\pi} \int_{-\infty}^{\infty} d k'
\frac{g(k')}{k-k'-i\eta} \ , \label{int_zeta}\\
&&\xi(k) =\frac{+i}{2\pi} \int_{-\infty}^{\infty} d k'
\frac{g(k')}{k-k'+i\eta} \label{int_xi} \ ,
\eea
where $\eta$ is an infinitesimal positive number.
>From the identity
\be
\frac{-1}{k-k'-i\eta}+\frac{1}{k-k'+i\eta} = -2\pi i\delta(k-k') 
\label{eqIIa4}
\ee
it follows that
\be
g(k)=\zeta(k)+\xi(k) \ .
\ee
Equations (\ref{g})-(\ref{int_xi}) can be written as well
in the following way:
\bea
&&\Psi(x,t)\equiv\int_{-\infty}^{\infty} dk g(k) \ \psi_k(x,t) ,
\label{psi} \\
&&\Phi(x,t)\equiv\int_{-\infty}^{\infty} dk g(k) \ \phi_k(x,t) ,
\label{phi} \\
&&R(x,t)\equiv\int_{-\infty}^{\infty} dk g(k) \ \rho_k(x,t)
\label{rho} ,
\eea
where
\bea
&& \psi_k(x,t) = \exp[i (kx-\omega_k t)] \ ,
\label{exp_psi} \\
&& \phi_k(x,t) =\frac{-i}{2\pi} \int_{-\infty}^{\infty} d \kappa \
\frac{\exp[i (\kappa x-\omega_{\kappa} t)]}{\kappa
-k-i\eta} \ ,
\label{exp_phi}\\
&&\rho_k(x,t) =\frac{+i}{2\pi} \int_{-\infty}^{\infty}
d \kappa
\frac{\exp[i (\kappa x-\omega_{\kappa}
t)]}{\kappa-k+i\eta} \label{exp_rho} \ , \\
&&\psi_k(x,t) = \phi_k(x,t) + \rho_k(x,t) \ .
\eea
For $t=0$ we obtain, as required by their definitions,
\bea
&& \phi_k(x,0) =\Theta(x) \psi_k(x,0) , \\
&& \rho_k(x,0) =\Theta(-x) \psi_k(x,0) .
\eea

We now invoke the 
premises of causality and superluminal
propagation, focusing 
on the ARS model involving
the dispersion relation
\be
\omega_k = c \sqrt{k^2 - m^2} \label{omega} \ .
\ee
As long as $|k|>m$ this dispersion relation
describes normal oscillating modes. Unstable
modes exist for $|k|<m$. One might attempt to
avoid the unstable modes altogether by choosing
a $g(k)$ that vanishes or is negligibly small for
$|k|<m$. This can be done, for example, by
choosing an initial state with a Gaussian $g(k)$,
centered around $k_o$ and having a width 
$\Delta k_o$ such that $|k_o \pm \Delta k_o|>>m $. This
corresponds in the case of the optical amplifier to
a pulse detuning large compared with a radiative decay
rate.
It turns out, however, as might be expected from the example
of the optical amplifier, that even for such an initial wave packet
$\Psi(x,0)$ the unstable modes play an
essential role in the time evolution of both the
truncated and the residual wavepackets,
$\Phi(x,t)$ and $ R(x,t)$, respectively.

Consider the integrals in
(\ref{exp_phi}) and
(\ref{exp_rho}) as contour integrals in the complex
$\kappa$ plane. The integrands, analytically
continued into the complex $\kappa$ plane, each have
a single, simple pole above or below the real
$\kappa$ axis at $\kappa=k\pm i\eta$, and both
have two branch points at $\kappa=\pm m$, which we
connect with a branch cut on the line
segment $(-m,m)$ on the real $\kappa$ axis. The
contour from $-\infty$ to $\infty$ should pass, as
usual, slightly above the real $\kappa$ axis (at a
distance smaller than $\eta$). As shown below,
this ensures causality according to equation
(\ref{eqII2}). In the limit of infinite
$|\kappa|$,
\be
\lim_{|\kappa|\rightarrow \infty} \omega_{\kappa}=
c \kappa ,
\ee
and on the circle at infinity,
\be \kappa x - \omega_{\kappa}t\longrightarrow
\kappa (x-c t) .
\ee
For $x>ct$ we can therefore close the contour
integral in the upper-half plane, whereas for
$x<ct$ we close the contour in the
lower half. In both cases the contributions to
the integral from the arcs at infinity vanish.

Using first the residue theorem for $x>ct$, we see
immediately that the superluminal parts of the
time-evolved residual and truncated wave packets
satisfy
\bea
&&\rho_k(x>c t,t) = 0 , \\
&&\phi_k(x>c t,t) =\exp[i(k x -
\omega_k t)] .
\eea
These results are not surprising, as
they simply reformulate equations
(\ref{eqII2}) and (\ref{eqII6}), respectively.
For $x<ct$, where we close the contour in the
lower half-plane, the integral encircles the branch
cut on $(-m,m)$. After deforming the contour
and isolating contributions from this branch cut, we
use the residue theorem and obtain
\bea
&&\rho_k(x<c t,t) = \exp[i(k x -\omega_k t)] \ - \
I_k^{\rho}(x,t),
\\ &&\phi_k(x<c t,t) = I_k^{\phi}(x,t),
\eea
where
\bea
&& I_k^{\rho}(x,t) =
\frac{i}{2\pi} \int_{C} d \kappa \
\frac{\exp[i (\kappa x- c t \sqrt{\kappa^2-m^2}
)]}{\kappa -k-i\eta} \ ,\\
&& I_k^{\phi}(x,t) = \frac{i}{2\pi} \int_{C} d
\kappa \
\frac{\exp[i (\kappa x- c t \sqrt{\kappa^2-m^2}
)]}{\kappa -k+i\eta} \ ,
\eea
and $\int_{C} d \kappa$ is a closed contour
circling counter-clockwise the branch cut on the
line segment $(-m,m)$ while not circling the poles
at $k\pm i\eta$.
Each of the integrals, $I_k^{\rho}(x,t)$ and $I_k^{\phi}(x,t)$,
is dominated by a saddle point on the imaginary $\kappa$ axis in
the complex $\kappa$ plane and exponentially grows with time. 
Combining terms, we obtain
\be
R(x,t) = \Theta (ct-x) \Psi(x,t) \ - \   \Theta
(ct-x) \int_{-\infty}^{\infty} dk \ g(k) \ I^\rho_k(x,t), 
\label{residual}
\ee
\be
\Phi(x,t)
= \Theta (x-ct) \Psi(x,t) \ + \   \Theta
(ct-x) \int_{-\infty}^{\infty} dk \ g(k) \ I^\phi_k(x,t).
\label{truncated}
\ee
The integrals give
exponentially growing contributions to the luminal parts of both the
truncated and residual wavepackets. Our choice of $g(k)$ enforces
$|k\pm i\eta|>m$, and as a result, 
\be
\int_{-\infty}^{\infty} dk \ g(k) \ \left[ \ I^\phi_k(x,t) - I^\rho_k(x,t)
\ \right] = 0
\ee
We see therefore that, 
when the residual and truncated
wavepackets (\ref{residual}) and (\ref{truncated}) are combined to form the 
complete wavepacket $\Psi(x,t)$ [Eq. (\ref{eqIIa3})], the 
exponentially growing luminal parts cancel each other. 

\subsection*{Discussion and implications for
quantum noise}

We are studying the time evolution of three
wavepackets: the complete wavepacket
$\Psi(x,t)$, the truncated wavepacket
$\Phi(x,t)$, and the retarded, residual
wavepacket $R(x,t)$.  These three wavepackets
can be decomposed in two different ways. In equations
(\ref{g})-(\ref{int_xi}) they were decomposed in
the usual way via a Fourier transform at the
initial time $t=0$ into normal and unstable modes.
The Fourier components of the truncated wavepacket
$\zeta(k)$ and the retarded wavepacket
$\xi(k)$ are related to the Fourier components of
the complete wave packet, $g(k)$, by equations
(\ref{int_zeta}) and (\ref{int_xi}), respectively.
If we choose to construct the complete wave
packet from normal modes $g(k)$, where $|k|\gg
m$, the truncated and retarded wavepackets will
have a strong unstable-mode component in them.
This was discussed by ARS, who
pointed out that because of the unitarity of the
time evolution, the unstable modes are
accompanied by an enhancement of the quantum noise.

In order to identify the noise in a space-time
picture we employed in equations
(\ref{psi})-(\ref{exp_rho}) a less common
decomposition. The difference between
(\ref{psi})-(\ref{exp_rho}) and
(\ref{g})-(\ref{int_xi})  lies in the order of integration.
Both decompositions can be derived from
\bea
&&\Psi(x,t)\equiv\int_{-\infty}^{\infty} dk g(k)
\exp[i (kx-\omega_k t)] , \nonumber \\
&&\Phi(x,t)\equiv
\frac{-i}{2\pi}\int_{-\infty}^{\infty} dq
\int_{-\infty}^{\infty} dp \
\frac{g(p) \exp[i(qx-\omega_q
t)]}{q-p-i\eta} \ , \\
&&R(x,t)\equiv
\frac{i}{2\pi}\int_{-\infty}^{\infty} dq
\int_{-\infty}^{\infty} dp \
\frac{g(p) \exp[i(qx-\omega_q
t)]}{q-p+i\eta} \ .
\eea
Equation (\ref{psi}) describes a
wavepacket made of a superposition of oscillating
waves
$\psi_k(x,t)\equiv \exp[i (kx-\omega_k t)]$, with
the momentum distribution $g(k)$. In
(\ref{phi}) and (\ref{rho}) each of these
oscillating waves is replaced by a new
wavefunction,
$\phi_k(x,t)$ and $\rho_k(x,t)$, respectively. The
weight function for the superposition forming the
respective wavepackets remains $g(k)$, but $k$ has
lost its meaning as a physical momentum. At any
time, $\psi_k=\phi_k+\rho_k$. At $t=0$, $\phi_k$
and $\rho_k$ are obtained from
$\psi_k$ by truncation. At a later time
$t>0$, one can distinguish between two regions. In
the superluminal region where $x>ct$,
$\phi_k =\psi_k$ and $\rho_k=0$. In the luminal
region where $x<ct$,
$\phi_k \neq \psi_k$: While $\psi_k$ is everywhere
a periodic wavefunction oscillating in space and
time,
$\phi_k$ is in this region 
exponentially growing as a function of
both $t$ and $x$; it is not oscillating in this
region. In the same retarded region $\rho_k$ has
a periodic oscillating component equal to $\psi_k$
and an exponentially growing component which
exactly cancels the contribution of $\phi_k$ to
this region.

The three wavepackets we consider are formed by
superpositions of these different wavefunctions
with the same weight function
$g(k)$. They evolve in time in the following way.
In the superluminal region $x>ct$ the
oscillating wave functions $\psi_k=\phi_k\equiv
\exp[i (kx-\omega_k t)]$, with $\omega_k$ given
above, combine to form a wavepacket moving at the
group velocity $v_g>c$; this is the superluminal signal.
In the luminal region $x<ct$ the
oscillating wave functions 
combine to cancel each
other. This cancellation 
ensures the unitary time evolution of the
complete wavepacket.
The residual part of the complete wavepacket is
essential for this cancellation to occur. 

Using the language of truncated wavepackets introduced
by ARS, we see that the superluminal signal is
constructed completely from the time evolution of
the forward tail, i.e., from the time evolution
of the truncated wavepacket. This truncated
wavepacket evolves with time into a
combination of the superluminal signal and an
additional, exponentially increasing part in the
luminal region $x<ct$. As discussed below, this 
additional part that grows exponentially with
time can be expected to be accompanied by substantial
quantum noise, as ARS observed using a different
decomposition of the same truncated wavepacket.

The new decomposition presented here therefore leads
us to conclude that the exponentially growing
 noise is {\it mostly} ``luminal" and
will be delayed compared with the superluminal
signal. This conclusion is consistent with the
exponentially growing noise due to SF in the case of the
optical amplifier \cite{remark2}.
Looking at the complete wavepacket, we
observe that contributions from the time-evolved
residual wavepacket will cancel in the luminal
region $x<ct$ the contributions from the
time-evolved truncated wavepacket. However, while
the signal in this region vanishes by the
cancellation of the two exponentially growing
contributions, {\it the noise does not cancel -- and
may be very large} \cite{analogy}.
We note that an amplification
of the signal in the superluminal region does
occur, but our new decomposition indicates that
this amplification is mostly a result of a rather
efficient constructive interference of oscillating
wavefunctions, while the luminal parts of the
time-evolved truncated and retarded wavepackets
appear to be controlled by the unstable modes. 

Our analysis in this section, being based on a ``first-quantization"
approach in which the wave packets are c-numbers, not operators,
has not dealt explicity with quantum noise. However, as in the theory
of the initiation of superfluorescence \cite{sf}, the linearity of the
model resulting from the approximation that there is no change in the
atomic inversion over the time scales of interest allows a treatment
of the operator fields as classical, fluctuating c-number 
fields \cite{bolda}. Thus the shaded
part of Figure 1a, the ``tail" from which the superluminal signal evolves,
becomes in such a treatment the truncated signal we have considered {\it plus}
a fluctuating noise field. In the limit of a very weak incident signal pulse,
the superluminal signal will be dominated by the noise part rather than 
the signal part of the tail shown in Figure 1a, and the signal-to-noise ratio
will therefore be small, consistent with the ARS results as
well as the results obtained 
in Section 4 for the model of an optical amplifier.

\section{Summary}
We have considered the effects of quantum noise on the propagation 
of a pulse with superluminal group velocity. In the case considered
by CKK \cite{chiao}, where an off-resonant, short pulse of duration $\tau_p$
propagates
with superluminal group velocity $v_g$ in an optical amplifier, we calculated a 
signal-to-noise ratio $SNR$ and found that, for an incident pulse consisting
of a single photon, $SNR<<1$ under the condition $(v_g/c-1)L>>\tau_p$
assumed by ARS \cite{ars} for discrimination between the pulse propagating in
the amplifier and a twin pulse propagating the same distance in vacuum. This 
result is fully consistent with the conclusions of ARS based on general
considerations and, in particular, the reconstruction of the superluminal
pulse from a truncated portion of the initial wave packet. However, if
we impose the weaker condition that $(v_g/c-1)L\stackrel{>}{\sim}\tau_p$,
then our conclusion is that $SNR>1$ is possible. However, in this case
superluminal group velocity is observable in the arrival statistics of
many photons, not per shot.

We showed that, in the case of the optical amplifier, the quantum noise is
attributable to the onset of superfluorescence, and could be associated
either with the quantum fluctuations of the field, along the lines of the ARS
considerations, or with the quantum fluctuations of the atomic dipoles. 

We then presented some general considerations based on unitarity
and causality and introduced a new wave packet decomposition. In
particular, we considered the ``residual" wave packet in addition to the
complete and truncated wave packets considered by ARS. This led to the
conclusion that the noise is mostly luminal, that in the luminal
region the truncated and residual signal grow exponentially but cancel
each other as required by unitarity, but that the {\it noise} is
not cancelled. For the case where the propagation time is large enough for
the superluminal signal to be clearly distinguished from a twin pulse 
propagated at the vacuum speed of light, our conclusions were again consistent
with ARS.
 
\section*{Acknowledgements}
We thank Y. Aharonov, E.L. Bolda, I.H. Deutsch, R.J. Glauber, P.G. Kwiat,
 B. Reznik, and A.M. Steinberg for helpful discussions or remarks during
 this work. This work was partially supported by the National Science
Foundation through a grant for the Institute for Theoretical Atomic
and Molecular Physics (ITAMP) at the Harvard-Smithsonian Center for
Astrophysics.
\section*{Appendix A}
The Heisenberg equations of motion for the Pauli operators follow from the
 Hamiltonian
(\ref{qq1}) and the commutation relations $[\sig_x,\sig_y]=2i\sig_z$,
etc.\ :
\be
\dot{\sig}_{xj}=-\om_o\sig_{yj} \ ,
\label{sigx1}
\ee
\be
\dot{\sig}_{yj}=\om_o\sig_{xj}+{2d\over\hbar}\sig_{zj}\hat{\calE}(z_j,t) \ ,
\label{sigy1}
\ee
\be
\dot{\sig}_{zj}=-{2d\over\hbar}\sig_{yj}\hat{\calE}(z_j,t) \ ,
\label{sigz1}
\ee
or, in the rotating--wave approximation,
\be
\dot{\sig}_j=-i\om_o\sig_j-{id\over\hbar}\sig_{zj}\hat{\calE}^{(+)}(z_j,t) \ ,
\label{sig1}
\ee
\be
\dot{\sig}_{zj}=-{2id\over\hbar}[\hat{\calE}^{(-)}(z_j,t)\sig_j-\sig_j^{\dag}
\calE^{(
+)}(z_j,t)] \ .
\ee

>From the formal solution of the Heisenberg equation of motion for 
$\hat{a}_k(t)$
 we obtain,
using equation (\ref{qq2}),
\bea
\hat{\calE}^{(+)}(z_j,t)&=&\hat{E}_o^{(+)}(z_j,t)+{2\pi id\over S\ell}
\sum_k\om_k\sum_{
i=1}^{N_T}
e^{ik(z_j-z_i)}\int_0^tdt'\sig_i(t')e^{i\om_k(t'-t)} \nonumber \\
&\equiv& \hat{E}_o^{(+)}(z_j,t)+\hat{E}_s^{(+)}(z_j,t) \ .
\eea
Here
\be
\hat{E}_o^{(+)}(z,t)=i\sum_k\left({2\pi\hbar\om_k\over S\ell}\right)^{1/2}
\hat{a}_k
(0)e^{-i\om_kt}
e^{ikz}
\ee
is the homogeneous (``vacuum") solution of the Maxwell equation
 for the quantized
field, while $\hat{E}_s^{(+)}(z,t)$ is the ``source" part. Now in
the mode
continuum limit $\sum_k\rightarrow (\ell/2\pi)\int dk=(\ell/2\pi c)\int
 d\om$,
\bea
\hat{E}_s^{(+)}(z_j,t)&=&{2\pi id\over S\ell}{\ell\over 2\pi
c}\sum_{i=1}^{N_T}\int_0^tdt'\sig_i(t')\int_{-\infty}^{\infty}d\om\om
e^{i\om(t'-t+[z_j-z_i]/c)} \nonumber \\
&=&-{2\pi d\over Sc}\sum_{i=1}^{N_T}\int_0^tdt'\dot{\sig}_i(t')\delta
(t'-t+[z_j-z_i]/c)
\nonumber \\
&\cong&{2\pi id\om_o\over
Sc}\sum_{i=1}^{N_T}\int_0^tdt'\sig_i(t')\delta(t'-t+[z_j-z_i]/c)
\nonumber \\
&=&{i\pi d\om_o\over Sc}\sig_j(t)+{2\pi id\om_o\over Sc}\sum_{i\neq
j}^{N_T}\sig_i(t-[z_j-z_i]/c)\theta(z_j-z_i)\theta(t-[z_j-z_i]/c) 
\nonumber  \\
&=&{i\pi d\om_o\over Sc}\sig_j(t)+\hat{E}'^{(+)}(z_j,t) \ .
\label{field1}
\eea
Here $\hat{E}'^{(+)}(z_j,t)$ denotes the field, at the position $z_j$ of 
atom $j$, that is
produced by all the {\it other} atoms of the medium.

We now use this result, and the operator identity $\sig_{zj}\sig_j(t)=-
\sig_j(t)$, in
equation (\ref{sig1}). The result is
\be
\dot{\sig}_j(t)=-i\om_o\sig_j(t)-\beta\sig_j(t)-{id\over\hbar}\sig_{zj}
(t)\hat{E}^{(+)}(z_j,t)
\ ,
\ee
where
\be
\beta={\pi d^2\om_o\over S\hbar c}
\ee
and $\hat{E}^{(+)}(z_j,t)=\hat{E}_o^{(+)}(z_j,t)+\hat{E}'^{(+)}(z_j,t)$.

Similarly, using the operator identity $\sig_j^{\dag}(t)\sig_j(t)={1\over
2}[1+\sig_{zj}(t)]$, we obtain from (\ref{sigz1}) and (\ref{field1})
\be
\dot{\sig}_{zj}(t)=-2\beta[1+\sig_{zj}(t)]-{2id\over\hbar}[\hat{E}^{(-)}(z_j,
t)\sig_j(t)
-\sig_j^{\dag}(t)\hat{E}^{(+)}(z_j,t)] \ .
\ee
Since the expectation value $\la\sig_z\ra$ of the TLA inversion operator
is $p_2-p_1 = 2p_2-1$, where $p_1$ and
$p_2$ are the lower-- and upper--state probabilities, respectively,
it follows that
$2\beta$ is the radiative (spontaneous emission) decay rate:
\be
{1\over\tau_{RAD}}=2\beta={2\pi d^2\om_o\over S\hbar c} \ .
\label{radd}
\ee
This is not the more familiar Einstein $A$ coefficient for spontaneous 
emission,
$A=4|{\bf d}|^2\om_o^3/3\hbar c^3$, because it gives the 
spontaneous emission rate
into modes propagating unidirectionally with a single polarization, whereas
$A$ is the spontaneous emission rate into all possible field modes in free
space. In fact $1/\tau_{RAD}$ is the spontaneous emission rate implicit in
much of laser theory: the coefficient $\lambda^2A/8\pi$ appearing in the
standard
expression for the gain coefficient $g(\nu)$, 
where $\lambda$ is the wavelength, is
just $1/\tau_{RAD}$ times the cross--sectional area $S$:
\be
g(\nu)={\lambda^2A\over 8\pi}(N_2-N_1){\cal L}(\nu)={1\over\tau_{RAD}}(N_2
-N_1)S{\cal L}(\nu) 
\label{radda}
\ee
for (nondegenerate) upper- and lower-level population 
densities $N_2$ and $N_1$,
respectively, an atomic lineshape function ${\cal L}(\nu)$, and
$|{\bf d}|^2/3=d^2$.

Finally we use the definitions (\ref{Fdef}) and (\ref{sdef}) of
$\hat{F}$ and $\hat{s}$ to obtain
\be
\hat{\dot{s}}_j(t)=-i(\Delta-i\beta)\hat{s}_j(t)-{id\over\hbar}
\sig_{zj}(t)\hat{F}(z_j,t),
\ee
\be
\dot{\sig}_{zj}(t)=-2\beta[1+\sig_{zj}(t)]-{2id\over\hbar}
[\hat{F}^{\dag}(z_j,t)
\hat{s}_j(t)-
\hat{s}_j^{\dag}(t)\hat{F}(z_j,t)]
\ee
in the rotating--wave approximation. The detuning between the TLA
resonance frequency
$\om_o$
and the central field frequency $\om$ is defined as $\Delta=\om_o-\om.$ 
Replacing
$\hat{s}_j(t)$ and $\sig_{zj}(t)$ by $\hat{s}(z_j,t)$ and $\sig_z(z_j,t)$, 
respectively, or
$\hat{s}(z,t)$ and $\sig_z(z,t)$ in the continuum limit, 
we obtain equations
(\ref{q4}) and (\ref{q5}).

\section*{Appendix B}
For the initial state in which all the TLAs are in the upper state,
$\la \hat{s}_i(t_o)\ra=0$ and $\la \hat{s}_i^{\dag}(t_o)
\hat{s}_j(t_o)\ra=\delta_{ij}$.
Then the operator
\be
\hat{{\cal S}}=\sum_{i=1}^{N_T}\hat{s}_i(t_o)
\ee
satisfies
\be
\la\hat{{\cal S}}\ra=0 \ ,
\label{b1}
\ee
\be
\la\hat{{\cal S}}^{\dag}\hat{{\cal S}}\ra=\sum_{i=1}^{N_T}\sum_{j=1}^{N_T}
\la
\hat{s}_i^{\dag}(t_o)\hat{s}_j(t_o)\ra=N_T \ .
\label{b2}
\ee
In the continuum limit
\be
\hat{{\cal S}}={N_T\over L}\int_0^Ldz\hat{s}(z,0) \ ,
\ee
\be
\la\hat{{\cal S}}^{\dag}\hat{{\cal S}}\ra={N_T^2\over
L^2}\int_0^Ldz'\int_0^Ldz''\la \hat{s}^{\dag}(z',t_o)\hat{s}(z'',t_o)\ra \ ,
\ee
and we can satisfy (\ref{b1}) and (\ref{b2}) by taking 
\be
\la \hat{s}(z,t_o)\ra = 0 \ ,
\label{cor1}
\ee
\be
\la \hat{s}^{\dag}(z',t_o)\hat{s}(z'',t_o)\ra={L\over N_T}\delta(z'-z'') \ .
\label{cor2}
\ee

\newpage
\noindent
{\bf Figure}\\ \\
\begin{figure}[htbp]
\epsfxsize=\textwidth \epsfbox{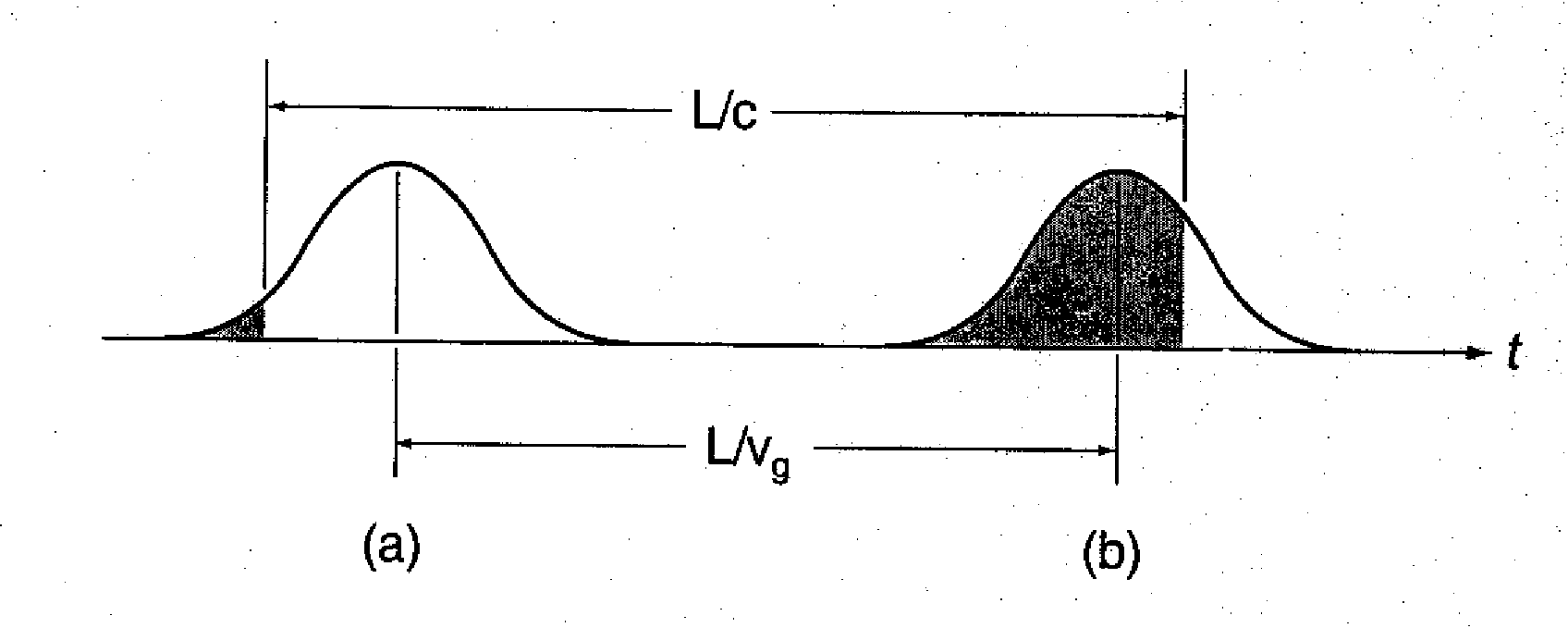}
\caption{  
Incident (a) and transmitted (b) signals for a propagation
length $L$ and group velocity $v_g>c$. It follows from the causal
connection between the two signals that the shaded portion of (b) is
completely determined by the shaded portion of (a). If $L(1/c-1/v_g)$
is much larger than the pulse duration, the peak of the transmitted
signal is reconstructed from a small tail of the incident pulse.}
\end{figure}


\begin{thebibliography}{99}
\bibitem{pi} The factor $1/\sqrt{\pi}$ in equation (\ref{qa4}) is a
consequence of our assumption of a Gaussian temporal profile for the
incident pulse.
For an arbitrary pulse shape $1/\sqrt{\pi}$ would be replaced by some
other factor less than unity.
\bibitem{remark2} The expression (\ref{p4}) for the refractive index is
the off-resonance approximation to the more general result
$n(\om)-1=(\om_p^2\Delta/4\om_o)/(\Delta^2+\beta^2)$, which gives
$v_g=c$ in the resonant case assumed in equation (\ref{stl}). Thus
the appropriate ``observation time" for SF is $L/c$.
\bibitem{remark} The fact that we obtain the noise term appearing in the
last line of (\ref{qa3}) after an integration of the radiatively broadened
gain profile over all
frequencies is consistent with the assumption made in obtaining (\ref{qa3}):
the observation time was assumed to be small compared with the radiative
decay time, implying a spectral bandwidth large compared with the 
gain bandwidth.
\bibitem{analogy} An analogy to a two-slit experiment can be made:
The enhancement of the signal on the screen at
places of constructive interference is consistent
with unitarity only if the destructive
interference at other places is considered as
well. The noise in the dark regions, however,
is not cancelled by any destructive interference
of the signals from the two slits.
\end{thebibliography}
\end{document}